# Thermal annealing of high dose P implantation in 4H-SiC

Cristiano Calabretta[1, 2]*, Massimo Zimbone[2], Eric G. Barbagiovanni[2], Simona Boninelli[2], Nico Piluso[3], Andrea Severino[3], Maria A. Di Stefano[3], Simona Lorenti[3], Lucia Calcagno[4] and Francesco La Via[2]

1) MIFT, Università degli studi di Messina, Viale F. Stagno d'Alcontres, 31 - 98166 Messina, Italy

2) IMM-CNR, VIII Strada, 5, 95121 Catania, Italy

3) STMicroelectronics, Stradale Primosole, 50, 95121 Catania, Italy

4) DFA, Università degli studi di Catania, Via S Sofia 64, 95123 Catania, Italy

*cristiano.calabretta@unime.it

**Keywords:** thermal annealing, crystal recovery, dopant activation, phosphorus, PL, Raman, TEM.

**Abstract**

In this work, we have studied the crystal defectiveness and doping activation subsequent to ion implantation and post-annealing by using various techniques including photoluminescence (PL), Raman spectroscopy and transmission electron microscopy (TEM). The aim of this work was to test the effectiveness of double step annealing to reduce the density of point defects generated during the annealing of a P implanted 4H-SiC epitaxial layer. The outcome of this work evidences that neither the first, 1 hour isochronal annealing at 1650 - 1700 - 1750 °C, nor the second one, at 1500 °C for times between 4 hour and 14 hour, were able to recover a satisfactory crystallinity of the sample and achieve dopant activations exceeding 1%.

**Introduction**

Powered semiconductor devices are key components for powered conversion systems. Silicon carbide (SiC) has received increasing attention as a wide-bandgap semiconductor suitable for high-voltage and low-loss power devices. Through new improvements in the crystal growth and SiC processing technology, the manufacturing of medium-voltage (600–1700 V) SiC Schottky barrier diodes (SBDs) and powered metal–oxide–semiconductor field-effect transistors (MOSFETs) has begun. However, dopant diffusion rates are slow even at temperatures as high as 1800 °C so that SiC cannot be doped effectively by thermal diffusion. Therefore, selected doped areas can only be created through ion implantation. However, the common drawback of ion implantation is the generation of lattice disorder. Typically, post-implantation annealing at high temperature (T > 1600 °C) is used to activate the dopant and decrease the defectiveness of the ion implanted area, thus increasing the crystal quality. The activation of the dopants in SiC is a very debated topic. Although in literature it is not rare to come across high dopant activations, these processes are strongly dependent on dopant solid solubility, which in phosphorus case is $4 \cdot 10^{18}$ at/cm$^3$ at 1750 °C in SiC [1] so that high dose implants result in low activations percentages after thermal equilibrium processes [2]. In this paper we will use Raman scattering and photoluminescence together with transmission electron microscopy to examine a series of P implanted 4H-SiC wafers that underwent a one step and a two step annealing processes and the doping activation from Raman spectra longitudinal optic mode will be extracted.

**Experimental Setup**

A 6 μm epitaxial layer was grown on (0001) 4° off axis 4H-SiC through low pressure hot wall chemical vapour deposition, and doped with N at a concentration of $10^{16}$ at/cm$^3$. P ion implantation was performed at 500 °C with energies between 20 and 200 KeV, and fluences ranging from $10^{13}$ to $10^{14}$ cm$^{-2}$ in order to obtain an almost uniform doped layer, 200 nm thick, with a concentration of about $10^{20}$ cm$^{-3}$. Following, isochronal (1 hour) thermal annealing were carried out at 1650 °C,

1700 °C, 1750 °C, followed by second thermal treatments at 1500 °C for times between 4 h and 14 h. During annealing processes graphitic capping layer was deposited on sample to avoid surface step bunching. Room temperature PL and Raman analyses were performed with a 325 nm He-Cd excitation source. Samples were characterized by a LabRAM HR Horiba Jobin Yvon spectrofluorimeter with a 1800 l/mm grating. The lattice structure of the implanted and annealed layers has been investigated by transmission electron microscopy (TEM) using JEOL JEM 2100 TEM-FEG operating at 200 kV.

**Results and discussion**

The PL spectra of implanted samples after annealing at different temperatures are reported in Fig. 1a. These spectra show two peaks, a narrow one related to the 4H-SiC bandgap at 390 nm and another related to defects induced by ion implantation peaked at 488 nm and with a 175 nm full width half maximum (FWHM). The presence of the defect peak is associated with crystallographic damage, which introduces optically active intra-bandgap energy levels [3]. The intensity of this peak increases with increasing the temperature of the first annealing process; in particular, at 1750 °C its intensity is 7% higher than at 1650°C. This behavior is mostly associated to a rise in the point defect density during the first annealing process; so that we can infer that the intra-bandgap energy level density grows with temperature.

Furthermore, the intensity of the PL peak changes after the second low temperature annealing, and in Fig. 1b this intensity is reported versus the temperature of the first annealing for different values of the second annealing time. The second annealing process at 1500 °C (for times between 4 h and 14 h) lowers the intra-bandgap 480 nm PL intensity peak in a range between 13% (1650 °C) to 20% (1750 °C) with respect to that underwent 1 h trial. This is due to a reduction in the carbon vacancies concentration $[V_c]$ in the implanted region [4]. Indeed, according to defect kinetics, during high temperature annealing, thermal generation of $V_c$ is the dominant process while at moderate temperature the recombination of $V_c$ with carbon interstitials present in the implanted area or injected from the graphitic capping layer appears to prevail. Moreover, this effect is accompanied by a slight enhancement of the band to band PL signal (Fig. 1c) which confirms a recovery of the crystalline structure.

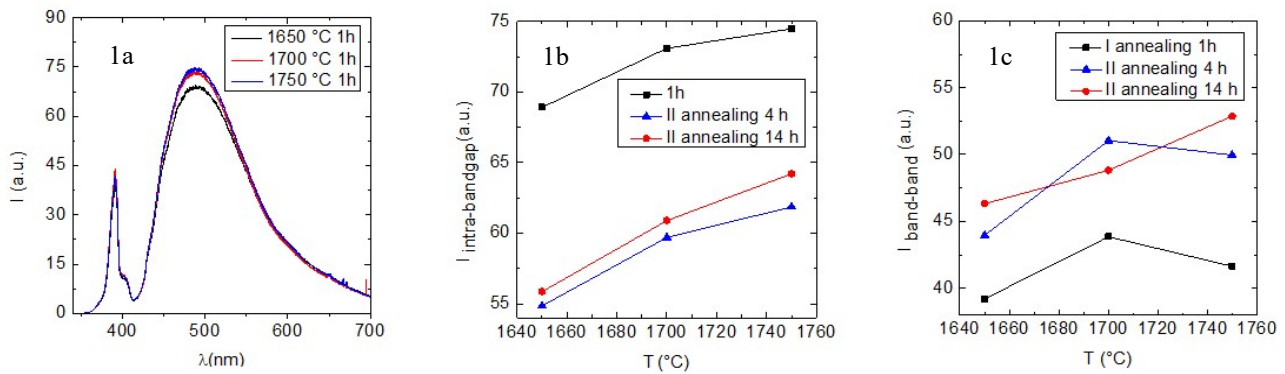

Fig.1.(a) PL spectra of the 1650 °C, 1700 °C and 1750 °C annealed P implanted samples, (b) PL intra-bandgap peak intensities, (c) PL band-band peak intensities for all annealed samples.

In Fig. 2a micro-Raman spectra from the three implanted samples after first HT annealing are reported: they show sharp Raman lines of tranverse optical (TO) modes at 776 cm$^{-1}$ and longitudinal optical mode (LO) at 964 cm$^{-1}$. In Fig. 2b the Raman width of the TO mode versus post-implantation annealing temperature is reported. The as grown sample TO curve shows a 4.4 ± 0.1 cm$^{-1}$ width, whose value after P implantation and annealing the TO width moves to 4.3 ± 0.1 cm$^{-1}$ and 5.8 ± 0.2 cm$^{-1}$ for 1650 °C and 1750 °C respectively.. Furthermore, the broadening of the TO peak in the 1750 °C annealed sample highlights the worsening crystal quality due to point defect creation in the lattice and is accompanied by a downshift in the TO mode frequency indicating an implantation induced tensile stress with respect to the supposed stress free as grown sample [5].

To test the effectiveness of the annealing procedure in terms of dopant activation, the LO mode was analyzed. In a polar semiconductor free electron collective excitations interacts with the longitudinal optical phonon via their macroscopic electric field leading to the LO phonon-plasmon coupled (LOPC) mode. The Raman line-shape of the LOPC mode varies considerably with the free carrier concentration, *n*. As reported in Fig. 2c the LO mode shows an asymmetric line-shape, which is a characteristic feature of the LO-phonon mode coupled with plasmons (LOPC). To evaluate the resulting carrier concentration, *n*, and mobility, *μ* arising from the different annealing processes we referred to the Raman scattering intensity formula [6] and extracted the carrier density and mobility values by fitting the LO peak of each annealed sample with the following relation (1)

$$I(\omega) = SA(\omega)Im\{-1/\varepsilon(\omega)\}, \tag{1}$$

where $\omega$ is the Raman shift, $S$ is a proportionality constant, $\varepsilon$ is the dielectric function and $A(\omega)$ is given in ref 4. The dielectric function $\varepsilon$ is given by

$$\varepsilon(\omega) = \varepsilon_\infty\{1 + [\omega_L^2 - \omega_T^2/(\omega_T^2 - \omega^2 - i\omega\Gamma)]\} - [\omega_p^2/\omega(\omega + i\gamma)] \tag{2}$$

and $\omega_p$ is the plasma frequency expressed by

$$\omega_p = \sqrt{4\pi n e^2/\varepsilon_\infty m} \tag{3}$$

where $m$ and $\varepsilon_\infty$ are the effective mass and the high frequency dielectric constant respectively. Mobility values are obtained from

$$\mu = e/m^*\gamma. \tag{4}$$

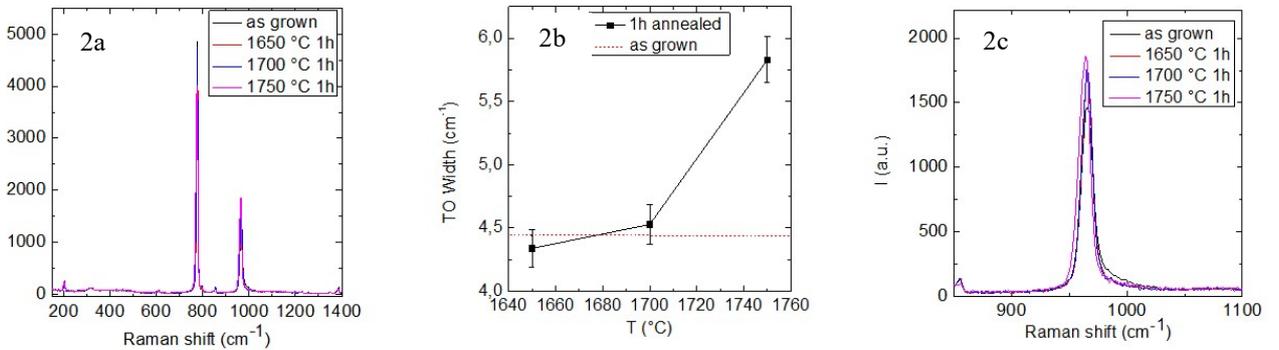

Fig. 2.(a) Raman spectra, (b) TO width and (c) LO Raman peaks for 1650 °C, 1700 °C and 1750 °C annealed P implanted samples.

The values of mobility and carrier concentration determined by the fit of Raman spectra in the samples that underwent a one hour annealing process are characterized by μ and n going from 68.5 ± 3 cm$^2$/Vs and 7.5 ± 0.4 ·10$^{17}$ e/cm$^3$ for 1650 °C annealing to 35.2 ± 3 cm$^2$/Vs and 6.73 ± 0.4 ·10$^{17}$ e/cm$^3$ for 1750 °C annealing, respectively.

The second annealing process produced only a slight increase in the dopant activation. The carrier concentration reaches a value of 9.9 ± 0.2 ·10$^{17}$ e/cm$^3$. Remarkably, the fitted mobility remained far from the expected values expressed by Caughey-Thomas equations [7] (Fig. 3a). Indeed, point defects behave as high density carrier scattering centers that heavily influence the mobility values.

High Resolution Transmission Electron Microscopy (HRTEM), shown in the inset of Fig. 3b, evidences that under the [11-20] zone axis an extra plane in the {0001} family is formed with a surrounding stress area leading to dislocation loop.

The persistence of an extended defect network of almost orthogonal traces made by extra-planes with dislocation loops (DLs) was also observed in the projected range of the implanted area Fig. 3b. Most of extra-planes are aligned parallel to the surface while a smaller percentage exhibits a perpendicular habit plane.

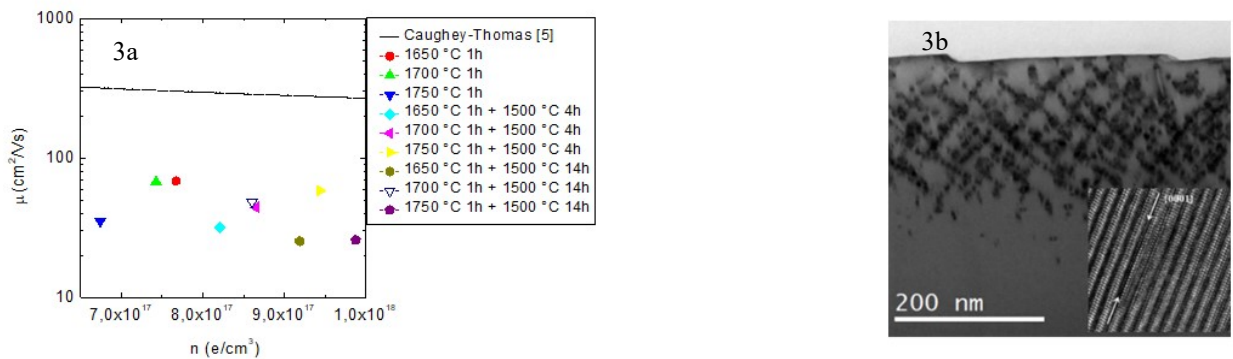

Fig. 3.(a) Electron mobility versus carrier concentration values extracted by the fitting procedure for the annealed samples, (b) XTEM image of 1750 °C 1 h + 1500 °C 14 h annealed sample. In the inset HRTEM image of an extra–plane in the {0001} plane family.

**Conclusions**

In this paper an analysis on P implanted 4H-SiC samples underwent double step annealing was presented. First annealing, in the temperature range 1650-1750 °C commonly used for dopant activation, is responsible for a wide intra bandgap PL signal. The samples evidence a tensile stress that widens and shifts the TO peak in the Raman spectra with respect to the as grown sample. Second annealing process (1500 °C) gives rise to a partial reduction of intra-bandgap signal, however it was not able to induce a significant increase of the dopant activation which remains below 1% of the total amount of implanted phosphorus. Moreover, carriers mobility remain far from the Caughey-Thomas equation.

**Acknowledgments**

This work was carried out in the framework of the ECSEL JU project WInSiC4AP (Wide Band Gap Innovative SiC for Advanced Power), grant agreement n. 737483.

**References**

[1] V. Simonka, A. Hossinger, S. Selberherr, and J. Weinbub. 123, (2018) 235701

[2] J. Senzaki, K. Fukuda, and K. Arai, J. Appl, Phys. 94, (2003) 2942;

[3] E. Fontana, N. Piluso, A. Russo, S, Lorenti, C.M. Marcellino, S. Coffa, F. La Via. Material Science Forum. 858 (2016) 418-421.

[4] H. M. Ayedh, R. Nipoti, A. Hallén, and B. G. Svensson, J. Appl. Phys. 122 (2017) 025701.

[5] S. Nakashima, T. Mitani, J. Senzaki, H. Okumura et al. J. Appl. Phys. 97 (2005) 123507.

[6] S. Nakashima and H Harima: Phys. Stat. Sol. 162 (1997) 39.

[7] T. Kimoto, and J. Cooper, Fundamental of Silicon Carbide, IEEE Wiley, 2014, pp. 23-24.